\documentclass[aps,prl,twocolumn,showpacs,preprintnumbers,superscriptaddress,amsmath,amssymb]{revtex4}
\usepackage{graphicx}
\usepackage{dcolumn}
\usepackage{bm}
\usepackage{ulem}
\usepackage{color}
\renewcommand{\v}[1]{{\bf #1}}

\newcommand{\s}{{\sigma}}

\def\be{\begin{eqnarray}}
\def\ee{\end{eqnarray}}

\newcommand{\nn}{\nonumber\\}

\newcommand{\Eq}[1]{Eq.~(\ref{#1})}

\newcommand{\Fig}[1]{Fig.~(\ref{#1})}

\begin{document}
\title{Topological insulators on a Mobius Strip}
\author{Langtao Huang}
\affiliation{Department of Physics, Tsinghua University, Beijing
100084, China;}
\author{Dung-Hai Lee}
\affiliation{ Department of Physics,University of California at
Berkeley, Berkeley, CA 94720, USA;} \affiliation{Materials Sciences
Division, Lawrence Berkeley National Laboratory, Berkeley, CA 94720,
USA.}

\date{\today}
\begin{abstract}
We study the two dimensional Chern insulator and spin Hall insulator
on a non-orientable Riemann surface, the Mobius strip, where the usual
bandstructure topological invariant is not defined. We show that while the flow pattern of edge
currents can detect the twist of the Mobius strip in the case of Chern insulator,
it can not do so in spin Hall insulator.
\end{abstract}
\maketitle

Band insulator with interesting bandstructure topology, the so called ``topological insulator'', has attracted considerable attention lately\cite{topo}. Mathematically these free-electron insulators are characterized by topological invariants in their bandstructure. Examples include the  Chern number\cite{TKNN,qi} and $Z_2$ index\cite{km,moore,roy,kane} for the Chern and $Z_2$ insulator, respectively. Physically a hallmark of these
insulators is their protected boundary states. In two dimensions these are the chiral edge states of the integer quantum Hall effect\cite{halperin,hatsugai}, and the helical edge states of the spin Hall insulator\cite{zb,konig}. In three dimensions the boundary states have a massless Dirac fermion dispersion relation\cite{fkm,arpes}.

In this paper we ask a simple question: in two dimensions can we put topological insulators on a non-orientable Riemann surface? and if so, what are their signatures? The first question was posed to us by Prof. X. Sun of the Fudan University. Clearly the usual bandstructure topological invariants can not be defined in this situation; therefore naturally one would examine the edge state structure. In the following we study two dimensional Chern and $Z_2$ insulator on the Mobius strip. Since the Mobius strip has only one edge, it is far from clear how does the flow pattern of the current look like.

First we start with the Chern insulator. As pointed out by Haldane\cite{Haldane_prl1988}, the necessary ingredient of Chern insulator is time-reversal symmetry breaking rather than net magnetic flux.
Specifically we consider the following two-band model on square lattice
\be
\mathcal{H}\left(\v k\right)=\v d(\v k)\cdot\vec{\tau}.\label{eq:Hk}
\ee
Here $\vec{\tau}$ are Pauli matrices which acts on the (two) orbital degrees of freedom, and
\be
\v d(\v k)=(\sin k_x,\sin k_y, 1-\cos k_x-\cos k_y).\ee It is straightforward to check that the
Chern number (or the TKKN index) associated with the valence band of this model is $1$. In order to implement this
model on the Mobius strip we first need to Fourier transform the above model to real space:
\begin{widetext}
\be
 H=-\frac{1}{2}\sum_{i,j}\left[i(\psi_{i + 1,j}^\dag\tau _x\psi_{i,j}+\psi _{i,j + 1}^\dag\tau _y\psi _{i,j})+\psi_{i+ 1,j}^\dag\tau_z\psi_{i,j}+\psi_{i,j + 1}^\dag\tau_z\psi_{i,j}+ h.c.\right]+\sum_{i,j}\psi _{i,j}^\dag\tau_z\psi _{i,j}.
\label{eq:HR}
\ee
\end{widetext}
In the above $(i,j)$ are the integer coordinates of the sites of a square lattice, and $\psi$ is a two-component fermion field associated with the two orbitals in question. Because we shall study \Eq{eq:HR} on the Mobius strip it is essential to define how do the orbitals couple to the local orientation.  A convenient definition is to let the pseudospin corresponding to the orbital degrees of freedom couple to curvature in the same way
the real spin in Dirac theory does\cite{lee}. This amounts to replacing the $\tau$ matrices in \Eq{eq:HR} by the following position-dependent Pauli matrices, i.e.,
\be
\tau_\mu\Rightarrow\vec{\tau}_{i,j;\mu}=\hat{n}_{i,j;\mu}\cdot\vec{\tau}.\label{localt}
\ee
Here $\mu=1,2,3$ and $\hat{n}_{ij,\mu}$ are unit vectors defining a local frame when the relevant surface is embedded in the three dimensional Euclidean space ($\hat{n}_3$ is the local surface normal).

\begin{figure}[tbp]
\begin{center}
\vspace{-0.4 in}
\includegraphics[scale=0.4]
{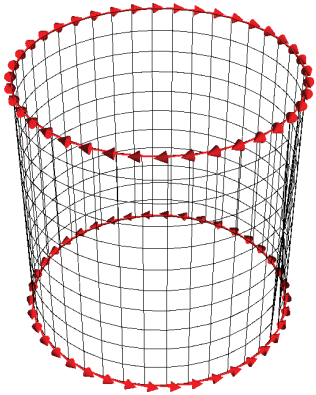}\vspace{-0.3 in}\caption{(color on-line) The edge current associated with the Chern insulator described in the text. There are 40/12 sites in the circumferential and height directions respectively. The direction of the current is illustrated by the red arrows. The length of arrows is proportional to the magnitude of the current. To make the plot easier to understand we have omitted the weak current away from the edge.\label{cylqhe}}
\end{center}
\end{figure}

Let us warm up by studying the surface of a cylinder. We build a coordinate system as follows:
\be
 x(u,v)=\cos u,~y(u,v)= \sin u,~z(u,v)=v.\ee
\noindent where $0\le u <2 \pi$ and $-1\le v\le 1$. Thus we have a cylinder of height 2 and radius 1. The local frame is defined by
\be
&&\hat{n}_1(u,v) =\partial _v\vec{r}(u,v)\nn
&&\hat{n}_2(u,v) =\partial _u\vec{r}(u,v)\nn
&&\hat{n}_3(u,v)=\hat{n}_1(u,v)\times\hat{n}_2(u,v),\label{frame}\ee where $\vec{r}(u,v)=(\cos u,\sin u,v)$. Set $u,v$ to a set of discrete values corresponding to the square lattice we substitute \Eq{frame} into \Eq{localt} then \Eq{eq:HR}. We diagonalize the resulting Hamiltonian numerically and compute the expectation value of the current
operator.
\be
j_{1,2}= \psi _{i,j}^\dag\tau _{i,j;1,2}\psi _{i,j}.\label{j}\ee
In \Fig{cylqhe} we plot the result. As expected counter-propagating chiral edge currents are found on the two opposite edges.

\begin{figure}[tbp]
\begin{center}
\vspace{-0.4 in}
\includegraphics[scale=0.4]
{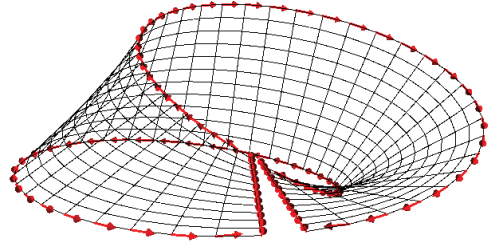}\vspace{-0.3 in}\caption{(color on-line) The edge current associated with the Chern insulator on a Mobius strip. There are $40\times 12$ sites, and all bonds across a line segment are removed. Again, to make the plot easier to understand we have omitted the weak current away from the edge.\label{mobiusqhe1}}
\end{center}
\end{figure}

\begin{figure}[tbp]
\begin{center}
\vspace{-0.4 in}
\includegraphics[scale=0.4]
{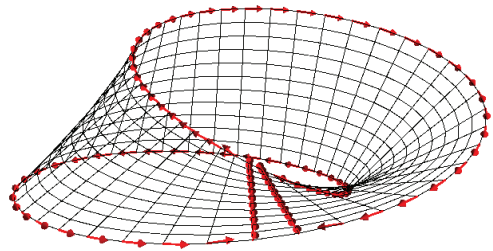}\vspace{-0.3 in}\caption{(color on-line) The edge current associated with the Chern insulator on a Mobius strip. There are $40\times 12$ sites. To make the plot easier to understand we have omitted the weak current away from the edge.\label{mobiusqhe2}}
\end{center}
\end{figure}

Now we are ready for the Mobius strip. The coordinate system we use is
\be
&&x(u,v)=\left(1 +\frac{v}{2}\cos \frac{u}{2}\right)\cos u\nn&&
y(u,v)=\left(1 +\frac{v}{2}\cos\frac{u}{2}\right)\sin u\nn&&
 z(u,v) =\frac{v}{2}\sin \frac{u}{2}.
 \label{eq:coordinate-m}
\ee
In \Fig{mobiusqhe1} we first present the result when all bonds across a line segment are removed.
We note that a pair of co-propagating edge currents are localized on the cut. Their presence is due to the
orientation flip across the cut. As a result the chirality of the Chern insulator reverses across the cut.
This is similar to the edge current produced at the location of magnetic field reversal in the quantum Hall effect.
Because the edge currents at the cut are {\it co-propagating}, seal the cut has no effect on them. The current pattern after the cut is sealed is shown in \Fig{mobiusqhe2}.
It is worthy to note that while geometrically there is no singularity on a Mobius strip, in defining the
chirality of the Chern insulator it is necessary to choose a cut across which the direction reverses. Thus the current
pattern of the Chern insulator does detect the twist of the Mobius strip.

\begin{figure}[tbp]
\begin{center}
\vspace{-0.4 in}
\includegraphics[scale=0.35]
{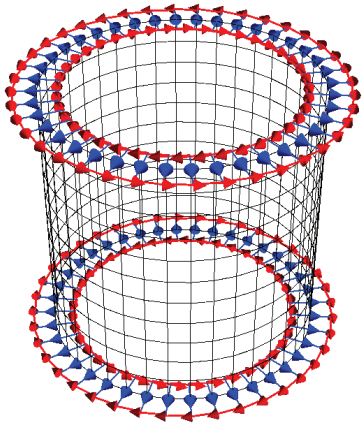}\vspace{-0.3 in}\caption{(color on-line) The edge current associated with the $Z_2$ insulator on a Mobius strip. The blue arrows indicate the direction of local $S_z$ axis.The red arrows at the top/bottom of the blue ones illustrate the current associated with $S_z=\pm 1$, respectively. There are $40\times 12$ sites, and to make the plot easier to understand we have omitted the weak current away from the edge.\label{cylshe}}
\end{center}
\end{figure}

Next we study the $Z_2$ insulator on the Mobius strip. The momentum space Hamiltonian in $\mathbb{R}^2$ is given by
\be
\mathcal{H}\left(\v k\right)=d_1(\v k)\tau_1\otimes I_2+d_2(\v k)\tau_2\otimes I_2+
d_3(\v k)\tau_3\otimes\s_3,\label{z2hk}
\ee
where $I_2$ is the $2\times 2$ identity matrix. The corresponding real space version is
\begin{widetext}
\be
 H&=&-\frac{1}{2}\sum_{i,j}\left[i(\psi_{i + 1,j}^\dag\tau _x\otimes I_2\psi_{i,j}+\psi _{i,j + 1}^\dag\tau _y\otimes I_2\psi _{i,j})+\psi_{i+ 1,j}^\dag\tau_z\otimes\s_z\psi_{i,j}+\psi_{i,j + 1}^\dag\tau_z\otimes\s_z\psi_{i,j}+ h.c.\right]\nn&+&\sum_{i,j}\psi _{i,j}^\dag\tau_z\otimes\s_z\psi _{i,j}.
\label{z2hr}
\ee
\end{widetext}
In defining \Eq{z2hr} on the Mobius strip we define the local spin Pauli matrices, $\sigma_{i,j;\mu}$, in the same way in \Eq{localt}. We also start by studying the edge current pattern on the cylinder. As shown in \Fig{cylshe} there are
a pair of time-reversal conjugate, counter-propagating edge currents at each edge. Here blue arrows indicate the direction of local $S_z$ axis. The red arrows at the top/bottom of the blue ones illustrate the current associated with $S_z=\pm 1$, respectively.

\begin{figure}[tbp]
\begin{center}
\vspace{-0.4 in}
\includegraphics[scale=0.35]
{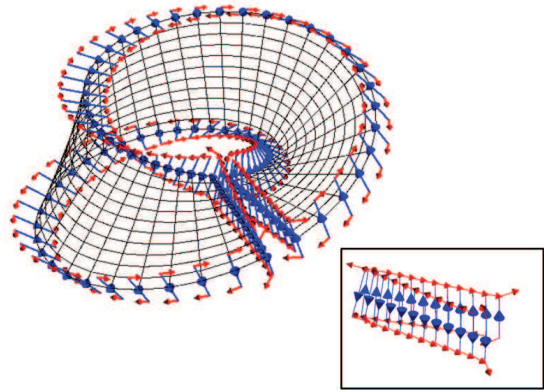}\vspace{-0.3 in}\caption{(color on-line) The edge
current associated with the $Z_2$ insulator on a Mobius strip. Bonds
across a line segment are removed. The meaning of blue and red
arrows are the same as in \Fig{cylshe}. There are $40\times 12$
sites, and to make the plot easier to understand we have omitted the
weak current away from the edge. The inset zooms in at the currents
near the cut. Note for each spin component there is a pair of
counter-propagating current.  \label{mobiusshe1}}
\end{center}
\end{figure}

\begin{figure}[tbp]
\begin{center}
\vspace{-0.4 in}
\includegraphics[scale=0.35]
{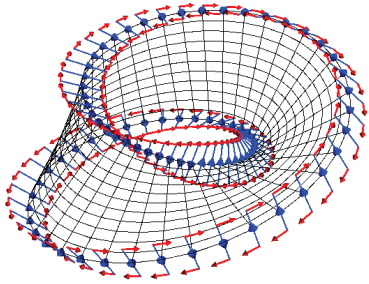}\vspace{-0.3 in}\caption{(color on-line) The edge current associated with the $Z_2$ insulator on a Mobius strip. The meaning of blue and red arrows are the same as in
\Fig{cylshe}. There are $40\times 12$ sites, and too make the plot easier to understand we have omitted the weak current away from the edge.  \label{mobiusshe2}}
\end{center}
\end{figure}

At last we study the $Z_2$ insulator on the Mobius strip. Again, we begin by removing the bonds across a line segment.
The associated current pattern is shown in \Fig{mobiusshe1}. The meaning of blue and red arrows are the same as in
\Fig{cylshe}. The inset zooms in at the currents near the cut.
It is important to note that for each spin component there is a pair of counter-propagating current. As the result, when the cut is sealed they are allowed to back scatter
against each other and hence gaps out the associated edge modes.
The result with the cut sealed is shown in \Fig{mobiusshe2}. As expected the edge currents associated with the cut are completely removed.

In summary we have studied the edge current distribution of the Chern and $Z_2$ insulators on the Mobius strip. The current pattern of the Chern insulator clearly detects the twist, while that of the $Z_2$ insulator does not. This reveals an interesting interaction between the topology of the electronic structure and that of the substrate. The fact that the $Z_2$ topological insulator can be seamlessly put on a $Z_2$ fiber bundle (the Mobius strip is a $Z_2$ fiber bundle over a circle) is particularly interesting. The mathematical meting of this needs to be clarified in the future.

{\bf Acknowledgment}: We thank Hong Yao for helpful discussions. LTH
is supported by the NSFC Grant No. 11074143, and the Program of
Basic Research Development of China Grant No. 2011CB921901. LTH also
acknowledges the support of China Scholarship Council and the
Doctoral Short-Term Visiting-Abroad Foundation of Tsinghua
University, Beijing. DHL is supported by DOE grant number
DE-AC02-05CH11231.

\end{document}